%% file: sigir.tex
\documentclass[sigconf]{acmart}

\usepackage{amsmath,amsfonts,amsthm,mathtools,latexsym}
\usepackage{rotating}
\usepackage{graphics}
\usepackage{graphicx}
\usepackage{booktabs}
\usepackage{multirow} 
\usepackage{url,bm,comment}
\usepackage{subcaption}
\usepackage{enumitem} 

\newtheorem{remark}{\bf Remark}

\makeatother

\newcommand{\bea}{\begin{eqnarray}}
\newcommand{\eea}{\end{eqnarray}}

\newcommand{\myeqref}{eq$.$~\eqref}

\newcommand{\E}{\mathbb{E}}

\newcommand{\pdfig}[3]{
\begin{figure}[htb]
\centering
\includegraphics[width=#1cm]{#2}
\caption{#3\label{fig:#2}}
\end{figure}
}

\newcommand{\Ac}{{\mathcal A}}

\newcommand{\Cc}{{\mathcal C}}

\newcommand{\Fc}{{\mathcal F}}

\newcommand{\Hc}{{\mathcal H}}
\newcommand{\Ic}{{\mathcal I}}

\newcommand{\Pc}{{\mathcal P}}

\newcommand{\Sc}{{\mathcal S}}
\newcommand{\Tc}{{\mathcal T}}

\AtBeginDocument{%
  \providecommand\BibTeX{{%
    Bib\TeX}}}

\copyrightyear{2025}
\acmYear{2025}
\setcopyright{cc}
\setcctype{by}
\setcctype{by}
\acmConference[SIGIR '25]{Proceedings of the 48th International ACM SIGIR
Conference on Research and Development in Information Retrieval}{July 13--18,
2025}{Padua, Italy}
\acmBooktitle{Proceedings of the 48th International ACM SIGIR Conference on
Research and Development in Information Retrieval (SIGIR '25), July 13--18,
2025, Padua, Italy}\acmDOI{10.1145/3726302.3730008}
\acmISBN{979-8-4007-1592-1/2025/07}

\begin{document}


\title{\!Information Retrieval in the Age of Generative AI: The RGB Model}


\author{Michele P. Garetto}
\affiliation{%
  \institution{University of Turin}
  \city{Torino}
  \country{Italy}}
\email{michele.garetto@unito.it}

\author{Alessandro P. Cornacchia}
\affiliation{%
\institution{ KAUST}
  \city{Thuwal}
  \country{Saudi Arabia}
}
\email{alessandro.cornacchia@kaust.edu.sa}

\author{Franco Galante}
\affiliation{%
  \institution{Politecnico di Torino}
  \city{Torino}
  \country{Italy}}
\email{franco.galante@polito.it}

\author{Emilio Leonardi}
\affiliation{%
  \institution{Politecnico di Torino}
  \city{Torino}
  \country{Italy}}
\email{emilio.leonardi@polito.it}

\author{Alessandro Nordio}
\affiliation{%
  \institution{Consiglio Nazionale delle Ricerche}
  \city{Torino}
  \country{Italy}}
\email{alessandro.nordio@cnr.it}

\author{Alberto P. Tarable}
\affiliation{%
  \institution{Consiglio Nazionale delle Ricerche}
  \city{Torino}
  \country{Italy}}
\email{alberto.tarable@cnr.it}

\renewcommand{\shortauthors}{Garetto et al.}

\begin{abstract}
The advent of Large Language Models (LLMs) and generative AI is fundamentally transforming information retrieval and processing on the Internet, bringing both great potential and significant concerns regarding content authenticity and reliability. This paper presents a novel quantitative approach to shed light on the complex information dynamics arising from the growing use of generative AI tools. Despite their significant impact on the digital ecosystem, these dynamics remain largely uncharted and poorly understood.
We propose a stochastic model to characterize the generation, indexing, and dissemination of information in response to new topics. This scenario particularly challenges current LLMs, which often rely on real-time Retrieval-Augmented Generation (RAG) techniques to overcome their static knowledge limitations. Our findings suggest that the rapid pace of generative AI adoption, combined with increasing user reliance, can outpace human verification, escalating the risk of inaccurate information proliferation across digital resources.
An in-depth analysis of Stack Exchange data confirms that high-quality answers inevitably require substantial time and human effort to emerge. This underscores the considerable risks associated with generating persuasive text in response to new questions and highlights the critical need for responsible development and deployment of future generative AI tools.
\end{abstract}

\ccsdesc[500]{Information systems~Web searching and information discovery}
\ccsdesc[500]{Information systems~Evaluation of retrieval results}
\ccsdesc[500]{Information systems~Novelty in information retrieval}
\ccsdesc[300]{Computing methodologies~Natural language generation}

\keywords{Web 
answering, Information quality, Large Language Models, \\Retrieval-Augmented Generation, Automation bias, Stack Exchange.  }





\maketitle

\input{intro}
\input{related}
\input{model_AT}

\input{scenarios}

\input{stackoverflow}

\section{Discussion and Conclusions}
To the best of our knowledge, ours is the first attempt to capture into a 
relatively simple analytical model high-level dynamics of topic-specific information 
across different digital resources, highlighting potential emergent phenomena and risks arising from evolving information retrieval behaviors. 
To this purpose, we had to strike a balance between simplicity and representational power.  

Our model incorporates parameters of various natures, which are difficult to determine from real data 
for several reasons:
\begin{enumerate}[leftmargin=0.6cm, topsep=1pt]

\item Some parameters are defined at a high aggregation level (e.g. the global rate of information on a topic indexed by search engines or included in training sets). These are difficult to obtain from Q\&A datasets that do not track global-scale dynamics.
\item Some are unknown (such as the rate and accuracy at which digital resources are crawled or fed into LLMs), as private companies keep their algorithms and metrics strictly confidential.
\item Some relate to the intrinsic quality of an answer, a metric that may be difficult to measure in the most general case.
\item Some concern human behavior during an ongoing paradigm shift in information retrieval, for which reliable measurements are either missing or very limited.
\item Most are changing rapidly, causing any dataset to become outdated within months of collection.
\end{enumerate}

For all the above reasons, trying to fit the parameters in Table \ref{tab1} using datasets makes little sense. The results produced by our model are not meaningful in absolute terms. They become interesting (in a relative sense) when we perform what-if analysis, i.e. when we vary a few crucial parameters while keeping the others fixed. We provided an example of what-if analysis in Sec. \ref{sec:scenarios}, where we compared two scenarios: GAI, describing answer dynamics for a hypothetical topic under current conditions, and post-GAI, describing dynamics of the same topic in an envisioned future state. 

The impressive increase in user requests to AI-powered assistants and the associated proliferation of AI-generated content on the web require a critical examination of future scenarios. To address this pressing issue, we have introduced an initial quantitative framework to assess and project the impact of generative AI tools on the information ecosystem. Our preliminary results suggest a significant risk of increased misinformation proliferation, driven by two key factors: the generation of persuasive or authoritative answers to topics that are not sufficiently consolidated, and the tendency of individuals to reduce their cognitive effort in evaluating different answers and discriminating information quality. We hope our research will make a meaningful contribution to the ongoing discourse on the responsible development 
of AI technologies.

\newpage

\bibliographystyle{ACM-Reference-Format}
\bibliography{sigir}

\end{document}

%% file: intro.tex
\section{Introduction} \label{sec:intro}

In recent years, the emergence of Generative Artificial Intelligence (GAI)
fueled by Large Language Models (LLMs) has greatly enhanced our abilities in retrieving information and interacting with digital content. The great success and widespread adoption of AI chatbots such as 
ChatGPT, Copilot, Gemini, Claude, Perplexity, Llama and many others are transforming both the 
way we search for information and the way we produce new digital content.

The rapid growth in data availability and advances in computing power have enabled LLMs to train on huge corpora, allowing them to generate text that is not only contextually relevant but also semantically rich. These capabilities have been used to enhance search engines and virtual assistants, providing users with more intuitive and ready-to-use answers to their queries.

While companies, researchers and individuals are striving to harness the potential of LLMs, serious concerns have also arisen about the potential for misuse and unintended consequences of these tools, such as the spread of misinformation and fake news.
Indeed, AI chatbots can generate responses that are syntactically and grammatically sound but factually incorrect.
Furthermore, LLMs can quickly produce coherent and persuasive text that tends to be presented as the definitive answer to the users.
This can increase the risk that users trust the information provided by these models without critically evaluating its accuracy or checking the original sources, thus fostering a culture of \textit{intellectual laziness}.
Also, the black-box nature of many LLM systems complicates efforts to ensure transparency and accountability in information dissemination.


Despite the vast amounts of information on which they are trained, LLMs face limitations when dealing with new or rapidly evolving topics. Since the training process relies on historical data, there is often a lag before the latest information is incorporated into the training datasets. This gap can make LLMs less reliable in answering questions about current events or emerging topics.
To mitigate these limitations, some LLMs are designed to integrate with traditional search engines or databases to access up-to-date information. This approach, commonly referred to as Retrieval-Augmented Generation (RAG), allows them to gather information from the internet in real-time and provide answers to questions that fall outside the scope of their training.
However, this integration is not always transparent to the user. When an LLM relies on external search engines, it may not clearly indicate that the answer comes from a real-time search and not from its internal knowledge.

Another insidious challenge 
in the use of LLMs is the risk of "autophagy" in training cycles. This term, borrowed from biology, refers to a process where AI-generated content becomes part of the training data for future models, creating a self-consuming loop. As LLMs produce vast quantities of text that are indistinguishable from human-written content, this generated content often finds its way back into the public domain. Over time, the inclusion of AI-generated content in training datasets could lead to a dilution of the originality and diversity of the generated responses, as models recycle their own interpretations rather than drawing from a diverse set of authentic, human-generated sources. 

The Web, as a primary source of information for billions, is especially vulnerable to the careless exploitation of emerging GAI technologies.
As we continue to integrate GAI tools into information retrieval and content creation, it is imperative to balance their innovative potential with responsible oversight to safeguard the integrity of information on the Internet.
It is thus crucial to develop proper methodologies and guidelines to
navigate this rapidly evolving landscape and better understand its future implications.

\subsection{Paper Contributions} \label{sec:contribs}
This paper proposes a novel analytical framework to describe and understand the new dynamics of information retrieval 
and dissemination triggered by the integration of GAI tools into daily workflows. 
To the best of our knowledge, our study presents the first-of-this-kind quantitative
approach to incorporate 
the several factors at play into a comprehensive framework capable of predicting the temporal trajectory 
of crucial key performance indicators. 

Specifically, we focus on the generation, retrieval, and replication of answers related to novel topics for which no prior information
is available --- a scenario that poses the greatest challenge and risk to any answer-generating system.
In this context, we examine the competition between conventional search engines and 
emerging generative AI systems employing hybrid strategies to generate answers,  
particularly real-time RAG.  Our analysis incorporates both 
algorithmic and human behavioral aspects. The primary objective is to forecast potential future scenarios if current trends in the adoption of generative AI tools persist.

Our analytical model is complemented by an in-depth examination of a large dataset from the Stack Exchange platform, containing questions and answers 
in computer science and mathematics. 
This investigation provides important insights into the intricate temporal dynamics of answers to novel and challenging questions. 
In particular, it highlights the considerable time often required for high-quality answers to surface.

%% file: related.tex
\vspace{-2mm}
\section{Related Work}
\label{sec:related}

Opportunities and risks of LLMs (more generally, foundation models) 
have been broadly discussed in \cite{rep22,gold23}.
In \cite{talk24} authors point out the potential for users to \lq anthropomorphize' 
GAI tools, leading to unrealistic expectations of human-level 
performance on tasks where in fact these tools cannot match humans.
The implications of LLMs being increasingly incorporated into 
scientific workflows are considered in \cite{binz23}.
Several studies have shown the risk for machine learning models to acquire factually incorrect knowledge during training and propagate 
incorrect information to generate content
\cite{parr21,kdd19,fact20,cao21,borji23}.
A framework to evaluate LLMs' ability to answer
complex questions is proposed in \cite{tan23}. 
While~\cite{vykopal24,zhou23,pan23} explore 
the disinformation capabilities of LLMs,
i.e., their ability to generate convincing outputs that
align with dangerous disinformation narratives.
The hallucination problem in natural language generation has been
a focus of several empirical studies \cite{ji23,zhang23,tonmoy24,niu24}.
Challenges posed by time-sensitive questions are explored
in \cite{chen21,kasai24,jia24}.
Techniques to detect generic LLM-generated text are presented in \cite{mit23,tang24},
while \cite{liu24,chern23} focus on methods to prevent and detect 
misinformation generated by LLMs.
A self-evaluation technique to enhance factuality is presented in \cite{self24},
while \cite{bai22} investigates the use of human feedback in LLM alignment.

Retrieval-Augmented Generation was first proposed in \cite{rag20} 
for generic knowledge-intensive NLP tasks.
Work \cite{wang24} proposes an adaptive mechanism to control access to external
knowledge by a conversational system. The use of search engine outputs
to increase LLM factuality is evaluated in \cite{fresh24}. 
The problem of explaining RAG-enhanced LLM outputs has been addressed
in studies such as \cite{menick22} and \cite{sudhi24}, highlighting the 
importance of interpretability in these hybrid systems.


The risk of autophagy in model training has received significant attention,
also from a theoretical point of view. 
The problem was first raised in the field 
of image processing \cite{autophagy, martinez2023, bertrand2024},
where feeding synthetic data back into models has been shown
to decrease both the quality and diversity of generated images.
Shumailov et al. 
considered this phenomenon for LLMs~\cite{collapse}~(earlier version of their work in~\cite{curse}), introducing the concept of \textit{model collapse}. 
They show that various models, when trained with data generated by previous generations of the same model, forget the tails of the original data distribution (early collapse) and tend to collapse into a distribution that is very different from the original one, typically with much lower variance (late collapse). 


Briesh et al.~\cite{self_consuming} 
consider three possible data sources for each generation of training data: the original dataset, fresh data, and data from any previous generation. 
Consistently across different combinations of these data sources, 
they observe a decrease in correctness in the first generation, which then increases across generations. However, this comes at the expense of diversity (especially for the model trained only with synthetic data), which collapses to zero. 

Dohmatob et al.~\cite{demystified} consider a (linear) regression task as a proxy for more complex models and characterize the test error when models are subsequently retrained with synthetic data. They find that performance decreases with the number of retraining generations. Gerstgrasser et al.~have extended this framework in~\cite{inevitable}, suggesting that data accumulation prevents model 
collapse for language models and deep generative models.

In \cite{chi24} authors have examined the correctness, consistency, comprehensiveness, and conciseness of ChatGPT answers to 517 programming questions on Stack Overflow, showing that 52\% of ChatGPT answers 
contain incorrect information, overlooked by users 39\% of the time. 
In ~\cite{public_good} authors show that activity on Stack Overflow has decreased 
since the advent of chat-like LLMs, as users rely more on generative AI tools, 
especially for coding questions. 

Lastly, we mention the work of Yang et al.~\cite{human_ai}, who
incorporate human behavior in self-consumption loops, similarly to us.
They show that AI-generated information tends to prevail in information filtering, whereas real human data is often suppressed, leading to a loss of 
information diversity.
 


%% file: model_AT.tex
\section{System Model}
\label{sec:model}
We model for simplicity a system comprising one conventional search engine
and one LLM integrated with it, using RAG to provide up-to-date answers to user queries.
This situation is common to many GAI systems: for example, ChatGPT/Copilot are tightly integrated with Bing, and the Gemini model is incorporated into Google's search infrastructure.  

\subsection{The RGB Model}
\subsubsection{Answer model for a given topic}
Consider a specific novel topic,  for which no prior knowledge is available anywhere in the system at time zero.
Similarly to the RGB color model, we assume that each potential answer to this topic is described by a length-3 vector,  which represents a convex combination of the three primary colors (red,  green and blue).
We emphasize that this is just for the sake of an intuitive, simple illustration of our model.
In general, we can have an arbitrary set $\mathcal{C} = \mathcal{C'} 
\cup \mathcal{C''}$ of primary colors, associated to different
initially generated answers, partitioned into a set $\mathcal{C'}$
of \lq good' answers and a set  $\mathcal{C''}$
of \lq bad' answers.
Although the model can be generalized to an arbitrary number of primary colors, three is the minimum required to distinguish between 'bad' (low-quality, biased, fake) answers, conventionally represented by red, and two distinct 'good' answers (with potentially different qualities), associated with blue and green. This allows us to quantify both the accuracy and diversity of the produced answers, as discussed in Section~\ref{sub:metrics}.
\footnote{In case we do not care about diversity of good answers, we could implement a simpler binary system with just two colors.}

The relative popularity (or strength) of answers of a given color in the system (or one of its subsystems) at time $t$ is determined by the number of \lq coupons' of the same color  in the considered subsystem.

We first consider the simpler case in which possible answers available in the system can only be red (1,0,0), green (0,1,0) or blue (0,0,1).  In Subsection~\ref{sub:ext}, we will present an extension of the model in which mixed colors (mixing some percentages of red, green, and blue) can be produced by either machine or human behavior.  

\pdfig{7}{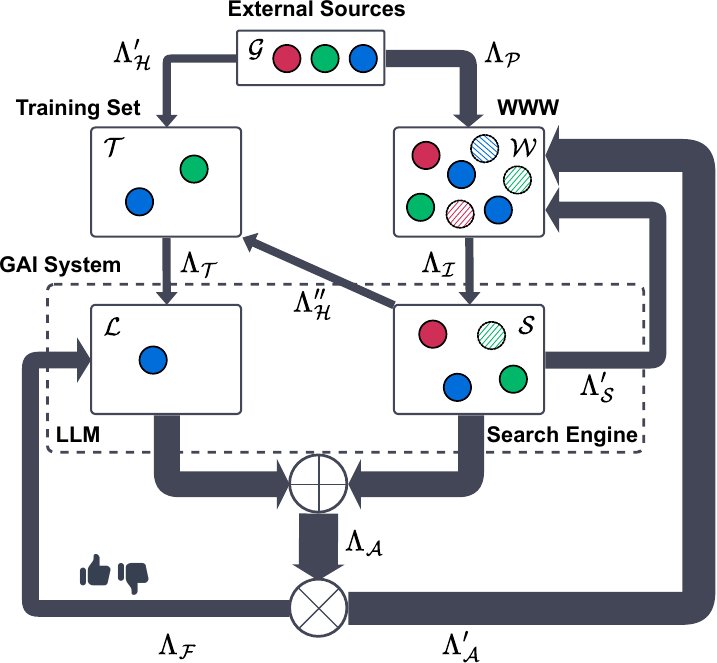}{
RGB model illustration: Colored circles represent coupons of answers to novel questions. Dashed circles are GAI-generated answers added to the Web. Gray arrows show generation/reinforcement/replication processes.
} 

\subsubsection{System compartments}
To specify how answers are initially generated,
and how they get replicated and reinforced across different digital resources over time, we introduce five subsystems (compartments)  
denoted by letters $G,W,T,S,L$ (see Figure \ref{fig:GPTmodel_new.pdf}).
At any given time $t \geq 0$, subsystem $s \in {\mathcal S} = \{G,W,T,S,L\}$
holds $N^s_c(t)$ \lq coupons' of color $c$,  $c \in \mathcal{C}$.  We denote by 
$N^s(t) = \sum_c N^s_c(t)$ the total number of coupons in subsystem 
$s$, at time $t$.          
The five considered compartments are:

\par \textbf{External sources} (G): this is a virtual place 
representing all information sources, such as news organizations, economists, political analysts, researchers, etc., from which answers are initially generated, i.e., introduced (through human intervention) into the digital ecosystem.

\par \textbf{Web} (W): this compartment represents the set of all online resources 
on the World Wide Web. 

\par \textbf{Training Set} (T): this is the ensemble of all
curated datasets used to train the LLM. 

\par \textbf{Search Engine} (S): the search engine continuously crawls the Web, indexing and ranking all found documents. It is then able to algorithmically assess and quantify the relevance of each indexed document in relation to a user's query. 

\par \textbf{Large Language Model} (L): the LLM constructs an internal representation of the information embedded within the training data, through a sophisticated architecture of interconnected parameters, or weights. Based on this representation, it produces a certain answer in response to a user's query.

\begin{remark}
For all compartments, the fraction of coupons of color $c$ is a measure of the strength of answer~$c$ compared to other answers. 
The \emph{absolute} number of coupons of color $c$ has a physical meaning
only for compartments W and T, where it represents the number 
of distinct replicas of answer $c$  found on the Web or in the Training Set,
respectively. For the Search Engine and the LLM, only the \emph{relative}
number of coupons of color $c$ matters, conveying a measure (weight) 
of how much that answer is considered relevant in the subsystem.
\end{remark}

\subsubsection{Coupon generation and propagation}

In our model, coupons accumulate in each subsystem due to generation, reinforcement, or replication processes (or flows), represented by arrows in Fig. \ref{fig:GPTmodel_new.pdf}. One exception is the external sources compartment,
which is supposed to contain a fixed number $N^G_c$ of coupons of color  $c$,
so that $g_c = N^G_c/N^G$ is the probability  of selecting uniformly at random a coupon of color $c$. This allows us to account for an arbitrary 
distribution of the color of fresh answers that are initially introduced
into the digital ecosystem.  
 	
We identify six main flows, indexed by $f \in \{\Pc,\Hc,
\Ic, \Tc, \Sc, \Ac\}$.  Each flow is modeled as a point process, characterized by either an aggregate rate $\Lambda_f$ or a per-coupon rate $\lambda_f$, 
depending on the physical meaning of the flow.
We will first introduce unbiased flow rates 
$\Lambda_f$ (or $\lambda_f$).  
The effective rate at which coupons are generated, reinforced, or replicated will be elucidated after the explication of our quality bias assumptions (Section \ref{subsec:bias}).    

\begin{description}[style=unboxed,leftmargin=0.5cm]
\item[$\Pc$]: this flow accounts for users posting fresh new answers on the Web, which is one of the ways answers are initially added to the digital information ecosystem. It is characterized by a (generally, time-varying) aggregate rate $\Lambda_\Pc(t)$.
\item[$\Hc$]: this is the overall process by which curated answers
are incorporated into the Training Set. This flow, characterized
by a relatively low aggregate rate $\Lambda_\Hc(t)$,  is divided into two sub-flows with rates 
$\Lambda'_\Hc(t) = \gamma \Lambda_\Hc(t)$ and 
$\Lambda''_\Hc(t) = (1-\gamma) \Lambda_\Hc(t)$,  where $0 \leq \gamma \leq 1$ 
is a model parameter.  The two  rates correspond respectively to fresh 
answers generated and directly incorporated into the Training Set 
(by-passing the Web), and to answers coming from the 
Search Engine, reflecting the fact that dataset curators often use 
the results of traditional search engines as sources of information.

\item[$\mathcal{I}$]: this process models how each answer posted on the Web is independently crawled and indexed by the Search Engine with per-coupon
rate $\lambda_\Ic(t)$. Note that the aggregate rate at which documents containing answers pertaining to the considered topic are indexed
is $\Lambda^\Ic(t) = N^W(t) \lambda_\Ic(t)$. 

\item[$\mathcal{T}$]:  this process models how each answer contained in the Training Set is fed into the LLM, with per-coupon rate $\lambda_\Tc(t)$, resulting
into an aggregate training rate $\Lambda_\Tc(t) = N^T(t) \lambda_\Tc(t)$ for answers
related to the considered topic.  


\item[$\Sc$]: This process represents users submitting queries related to the reference topic to the conventional search engine, with aggregate rate $\Lambda_\Sc(t)$.  Some of the answers obtained from the Search Engine are incorporated into documents posted again on the Web, creating a first
recursive loop in the model.  We denote the rate of answers 
fed back into the Web by $\Lambda'_\Sc(t)$.

\item[$\Ac$]: This process represents users submitting queries related to the reference topic to the GAI system combining the LLM with the Search Engine,   
with aggregate rate $\Lambda_\Ac(t)$.  Some of the answers obtained from the GAI system are fed back into the Web, at rate $\Lambda'_\Ac(t) = \alpha \Lambda_\Ac(t)$, 
$0 \leq \alpha \leq 1$, producing a second loop. 
Since many LLMs exploit the feedback provided by users 
(for example in the form of likes/dislikes) to further tune themselves,
it results into another feedback loop of rate $\Lambda_\Fc(t) = \beta \Lambda_\Ac(t)$,  which is relevant even if $\beta$ is small, because $\Lambda_\Ac(t)$
is large and rapidly growing.

\end{description}

The hybrid strategy (RAG) employed by the GAI system to respond to queries
is modeled by combining, for each color, the coupons contained in the LLM with those 
contained in the Search Engine: $N_c^A = N_c^L + N_c^S$. This way, the GAI system
is able to respond to queries even if the LLM has not yet been trained
with information related to the novel topic, provided that at least 
one answer is returned by the Search Engine.
However, the GAI system might prefer information derived from the LLM knowledge base
over that obtained from the Search Engine, when both are available.
Since a strict priority rule could be potentially dangerous in this context, we
consider the following soft preference mechanism: when an answer is incorporated into the LLM during the training process, we suppose that $w \geq 1$ coupons of the associated color are added to the LLM, rather than a single one. 
This approach permits modeling a generalized bias toward information stored 
in the LLM, without entirely disregarding external sources.

\subsubsection{Bias to quality}\label{subsec:bias}

We introduce, for each flow, a bias parameter towards the actual addition of coupons within the target subsystem of the flow. Such biases are introduced to take into account the effectiveness of algorithms/humans in promoting high-quality answers. 
We assume that each answer associated with a primary color component (three in the case of the RGB model) is characterized by an 
intrinsic quality $q_c$, normalized in such a way that $\sum_c q_c = 1$. 
Informally, we define the quality of each answer 
as the relative merit that would be attributed to it
by a large number of independent experts having unlimited time
to carefully evaluate and compare the answers.  Intrinsic qualities
are unknown to all actors in the system, both humans and machines. 
This fact is especially crucial for novel topics, for which 
none or insufficient effort has been made to evaluate proposed 
answers. Also, different actors have different willingness/ability
to identify and promote quality among answers.

To model this, we assume that each flow $f\in \{\Pc,\Hc,
\Ic, \Tc, \Sc, \Ac\}$, 
is characterized by a bias parameter $C_f$ towards quality.
Each time the flow is supposed to add a coupon of 
color $c$ in the target subsystem $s$, 
we assume that this coupon addition occurs only with probability: 
\begin{equation}\label{eq:rfi}
r^f_c= \frac{q_c + C_f}{\sum_{i\in \Cc} (q_{i} + C_f)}
\end{equation}
Note that $C_f = 0$ corresponds to the case in which coupon addition
occurs proportionally to intrinsic quality. As $C_f \rightarrow \infty$, coupon addition becomes oblivious to quality,  all answers
being treated the same. One can also consider negative values of $C_f$ (provided that $\min_c (q_c+C_f ) \geq 0$), 
representing the intention to further penalize low-quality answers in favor of high-quality answers.    

To summarize, at the times dictated by the point process, with aggregate rate
$\Lambda_f(t)$, associated with flow $f$, going from subsystem $s'$ to subsystem $s$,
a coupon is chosen uniformly at random among all coupons
currently stored in $s'$. Let $c$ be the color of the chosen coupon.  
A coupon of color $c$ is then added to the coupon stored in $s$ 
with probability $r^f_c$. 

\vspace{-2mm} 
\subsubsection{Finite coupon time-to-live}
In our model coupons do not stay forever in the subsystem in which they appear. 
With the exception of external sources, all compartments subject coupons to a finite duration of existence. The lifespan of each coupon in subsystem $s$ follows an exponential distribution with a rate parameter $\mu_s > 0$, representing the inverse of the expected time-to-live.
This stochastic removal process serves dual purposes:
For the Web and Training Set compartments, it simulates the natural obsolescence of information and subsequent removal by document maintainers.
In the Search Engine and LLM compartments, this mechanism reflects the 
necessity for periodic refreshment and validation 
for an answer to be considered still relevant.

\vspace{-2mm} 
\subsection{Model Solution}
Readers familiar with Markov Chains and queuing network will recognize
that our system can be described as a multi-class, open
system of four 
interconnected queues of type $\cdot/M/\infty$ associated to compartments $T,W,L,S$, fed by state-dependent external arrival processes
of \lq customers'. By construction, the giant Markov Chain representing the overall system state $\{N^s_c\}_{c,s}$ is ergodic, for any positive
values of arrival rates $\Lambda_f(t)$ ($f \in \{\Pc,\Hc\}$) 
and $\lambda_{f'}(t)$ ($f' \in \{\Ic,\Tc\}$).
Unfortunately, due to the complex replication process of coupons in the system,
our system does not allow a product-form solution for the stationary distribution of $\{N^s_c\}_{c,s}$.
To solve the model, one can resort to discrete-event simulation.
We implemented an ad-hoc simulator consisting of a simple C file \cite{repo}. 

We can also take a mean-field approach, and approximate the evolution
of the mean number of coupon $n^s_c(t) = \E[N^s_c(t)]$ of each color
in each compartment. Readers familiar with epidemic models such as the SIR model will recognize that our system
is similar to compartmental models used in epidemiology.
Following this approach, the system dynamics can be 
described by a set of ODEs (Ordinary Differential Equations).
We define for convenience $n^s(t) = \sum_c n^s_c(t)$ 
and $q^s_c(t) = \frac{n^s_c(t)}{n^s(t)}$.
We obtain the following system of coupled ODEs, which can be efficiently solved numerically. 
\begin{align}\begin{split}\label{eq:dyn_sys}
\dot n^T_c(t) & = \Lambda_\Hc\, r^\Hc_c [\gamma g_c + (1-\gamma) q^S_c(t)] \!- \mu_T\,  n^T_c(t) \\
\dot n^W_c(t) & = \Lambda_\Pc \, r^\Pc_c g_c \!- \mu_W\, n^W_c(t) +  q^S_c(t) \Lambda_\Sc'  \, r^\Sc_c +\!  q^A_c(t) \Lambda_\Ac' \,  r^\Ac_c \\
\dot n^L_i(t) & = n^T_c(t) \,w \, \lambda_\Tc(t) -\mu_L\,  n^L_c(t)  + q^A_c(t) \Lambda_\Fc(t)  \, r^\Fc_c  \\
\dot n^S_c(t) & = n^W_c(t) \lambda_\Ic(t)  \, r^\Ic_c - \mu_S\, n^\Sc_c(t)
\end{split}
\vspace{-0mm}
\end{align}
One technical issue
arising with this approach is that since $n^S(0) = 0$ (there are initially 
no coupon in the Search Engine), 
we have indeterminate $0/0$ forms for $q^S_c(0)$
and $q^A_c(0)$. We solved this problem assuming that
the Search Engine contains, initially, a given positive number $n^\Sc_b(0)$ of \lq black' coupons which, if chosen, do not produce any effect. Black coupons are never replenished, thus they gradually disappear from the system at rate $\mu_S$: $\dot n^S_b(t) = - \mu_S\, n^\Sc_b(t)$. Besides solving the above problem, black coupons serve another  purpose, i.e., they can model an answering machine (either the Search Engine or the GAI) preferring not to provide any answer (which happens when a black coupon is selected) in the initial phase in which none or very few colored coupons have been acquired: by setting $n^\Sc_b(0)$, one can
can thus model more or less \lq prudent' answering strategies
in the case of novel topics. A clear trade-off arises here
between limiting the dissemination of insufficiently consolidated answers (thereby mitigating the potential spread of misinformation), 
and maintaining user engagement and satisfaction.

Note that the above ODEs yield deterministic trajectories representing the \lq average' system evolution. They cannot be used to assess the variability of performance metrics, especially in the  
crucial initial transient phase. 
To obtain a more comprehensive understanding of system dynamics, 
and to characterize the distribution of possible 
trajectories, we must resort to simulating multiple independent 
runs of the system. 

\subsection{Metrics}\label{sub:metrics}
Several interesting metrics can be computed by solving our model over time. 
The fraction $\pi^s(t)$ of relevant answers contained in subsystem $s$, 
at time $t$, is:
$$\pi^s(t) = \frac{\sum_{c \in \mathcal{C}'} N_c^s(t)}{N^s(t)}$$
Note that $\pi^S(t)$ and $\pi^A(t)$ provide the {\em accuracy} of individual pieces of information generated, respectively,  by the Search Engine and the GAI system, at time $t$, pertaining to the considered topic.

Restricting to the set $\mathcal{C}'$ of relevant answers, one might also be interested in assessing the {\em diversity} degree of answers stored in subsystem $s$. Ideally, it would be desirable to achieve, for each relevant answer $c$, a fraction $p_c$ proportional to 
its intrinsic quality $q_c$ (renormalized among relevant answers): 
$$p_c = \frac{q_c}{\sum_{c' \in \mathcal{C}'} q_{c'}}, \qquad c \in \mathcal{C}'$$
If we instead obtain, in subsystem $s$ at time $t$, a fraction 
$$\hat{p}_c^s = \frac{N_c^s}{\sum_{c' \in \mathcal{C}'} N_{c'}^s}, \qquad c \in \mathcal{C}'$$
of relevant answers $c$, among all relevant answers contained in $s$, we could
quantify the {\em diversity} degree of $s$ by some distance metric between the discrete distributions $\{p_c\}_c$ and $\{\hat{p}_c^s\}_c$. For example, adopting the total variation distance, we can simply define the {\em diversity} degree $\rho^s(t) \in [0,1]$ (the higher the better):      
\begin{equation}\label{eq:rho}
\rho^s(t) = 1 - \frac{1}{2} \sum_{c' \in \mathcal{C}'} |p_{c'} - \hat{p}_{c'}^s(t)|
\end{equation}

Besides the accuracy and diversity of the subsystems,
we are especially interested in the following key performance indicators:

\par {\bf FIUA}: Fraction of Irrelevant Used Answers.
We call \lq used' the answers added to the set of online resources on the Web,
after having been suggested either by the Search Engine or by the 
GAI system. Note that we are not considering here answers
found on the Web which have been initially generated (by flow $\mathcal{P}$),
but only those produced by flows $\mathcal{S'}$ and $\mathcal{A'}$.
FIUA is the fraction of irrelevant used answers found on the Web, with respect to all used answers found on the Web (at any given time $t$).

\par {\bf AIRI}: AI Responsability Index.
Restricting to the set of irrelevant used answers found on the Web (at any given time $t$),
AIRI provides the fraction of them generated by
the GAI system (i.e., over flow $\mathcal{A'}$),
distinguishing them from those coming from the Search Engine (i.e., over flow $\mathcal{S'}$).

\par {\bf FRQ}: Fraction of Responded Queries.
Recall that our model incorporates a mechanism to simulate cautious answering strategies by initializing the Search Engine with a tunable number of black coupons: when a black answer is selected 
it is assumed that no usable response is returned to the user. 
Consequently, we can calculate, up to any given time $t$, 
the Fraction of Responded Queries (FRQ). Note that we can independently 
enable this mechanism for the Search Engine and/or the GAI system. 

\par {\bf AIAI}: AI Autophagy Index.
Our system contains several cycles over which information about the novel topic
can loop and reinforce itself. We are especially interested
in quantifying the autophagy of the LLM, i.e., the amount 
of information (about the topic) stored in it, which has been generated by the 
GAI system in response to users' queries. 
At any time $t$, AIAI is the fraction of coupons
found in the LLM, which have previously traversed 
flow $\mathcal{A'}$.  

\subsection{Model Extensions}\label{sub:ext}
The RGB model described so far can be extended to account for the generation
of coupons of intermediate colors, blending primary components 
in varying proportions.
  This process mirrors a common phenomenon observed in both human cognition and machine learning, wherein new digital content -- subsequently added to online resources on the Web -- is obtained by synthesizing information derived from multiple sources.     
  This aspect of information synthesis and propagation potentially represents a critical element that warrants incorporation into the model. We do so in two different ways, distinguishing human and algorithmic effects. In both cases we assume that
  just {two } sources are used to synthesize a new content posted 
  on the Web, but this assumption could be relaxed as well.
  
  For the GAI system, we assume that each generated response
  is derived from a linear combination of the two sources. 
  Specifically, a random fraction $u$ of the first source is combined with $(1-u)$ of the second, for each component, 
  where $u$ follows a uniform distribution over 
  the interval $[0,1]$. This reflects the fact that the GAI system is 
  not really able to prefer one source over the other (but note
  that sources to be mixed are first chosen proportionally to their 
  relative strength/popularity). A unique randomly generated text
  is then returned to the user, who will use it \lq as it is'.    
  
  We posit a different model for humans creating new content
  from the results returned by the Search Engine. Specifically,
  we propose that users combine a fraction $\xi$ of the perceived \lq best' source\footnote{The quality of a generic answer $i$ with components $\{x_c^i\}_c$ is defined as the weighted average of the intrinsic qualities of its components: $q_i = \sum_c x_c^i q_c$.} with a complementary fraction $(1-\xi)$ of the alternative source. The parameter $\xi$ is constrained to the interval $[1/2, 1]$, reflecting users' tendency to favor the perceived 
  trusted source. This model encapsulates the cognitive process by which users evaluate and discriminate between the relative quality of multiple alternatives. This process demands significant human effort, which is driving more and more people to delegate the task to artificial systems.

%% file: scenarios.tex
\section{Scenarios} \label{sec:scenarios}
Our model has a rich set of parameters, reflecting the utility of a versatile analytical framework for exploring the dynamics of a variety of possible scenarios. As a side effect, an exhaustive exploration of individual parameter impacts exceeds the scope of the present study. 
We therefore focus on select scenarios, reserving a more extensive sensitivity analysis for a forthcoming journal publication.

To demonstrate the insights that our model can provide,
we focus on a fixed set of parameters, and adopt an evolutionary 
perspective, by which just a few parameters are shifted to reflect 
current trends in technological adoption. Specifically, we propose the 
following three scenarios: 

\par {\bf pre-GAI}:
This scenario represents the conventional paradigm of information retrieval on the World Wide Web, before the availability of GAI services.
\par {\bf GAI}:
   This scenario aims to capture present conditions, 
characterized by rapid yet judicious adoption of GAI tools. It reflects a transitional period where users maintain vigilant oversight of GAI outputs while still predominantly relying on results generated by traditional search engines. 
\par {\bf post-GAI}:
   This prospective scenario envisions a future state where 
   information retrieval will be largely based on GAI services.
  It postulates a significant shift in user behavior, marked by increased confidence in and reliance on GAI-generated outputs, and 
a concomitant reduction of personal 
synthesis of information by employing search engines.

We consider an RGB model where primary components have intrinsic qualities $q_\mathrm{blue} = 0.5$, $q_\mathrm{green} = 0.4$, $q_\mathrm{red} = 0.1$. The set of \lq good' answers is $\mathcal{C'} = \{\mathrm{blue},\mathrm{green}\}$, while the red answer is bad, 
$\mathcal{C''} = \{\mathrm{red}\}$. All primary components are equally likely to be initially generated, so $g_c = 1/3$, $\forall c$.

Table \ref{tab1} reports the flow parameters chosen for the
{\bf GAI} scenario. The time unit of our system is the day, 
and we assume that all parameters remain constant over time\footnote{In actuality most parameters should be considered time-dependent due to the rapid evolution of information retrieval practices, 
but we adopt a simplifying assumption of constancy for the duration of the topic's lifespan, i.e., the temporal window in which the bulk of 
questions and answers related to the topic are generated.}.
So, for example, the first row in Table \ref{tab1}, which specifies
the posting process $\mathcal{P}$ of new answers on the Web, 
indicates that, starting at time 0, answers are posted with rate
of 1 per day, with a quality bias parameter $C_\mathcal{P} = 1$,
and they persist online for a duration that follows an exponential distribution with a mean of 100 days.

\begin{table}[tbhp]
\centering
\scalebox{0.85}{
\begin{tabular}{c|c|c|c|c} 
 \toprule
 \rule{0pt}{2ex} \textbf{flow} & \textbf{rate} & \textbf{bias} & TTL & related \\ 
 \midrule
 \rule{0pt}{2ex} $\mathcal{P}$ & $\Lambda_\mathcal{P} = 1$ & $C_\mathcal{P} = 1$ & $1/\mu_W = 100$ &   \\
 \rule{0pt}{2ex} $\mathcal{H}$ & $\Lambda_\mathcal{H} = 0.1$ & $C_\mathcal{H} = -0.08$ & $1/\mu_T = 1000$ & $\gamma = 0.5$    \\
 \rule{0pt}{2ex} $\mathcal{I}$ & $\lambda_\mathcal{I} = 0.1$ & $C_\mathcal{I} = 0$ & $\mu_S = \lambda_\mathcal{I}$ &  \\
 \rule{0pt}{2ex} $\mathcal{T}$ & $\lambda_\mathcal{T} = 0.1$ & -- & $\mu_L = \lambda_\mathcal{T}$ & $w = 3$  \\
  \rule{0pt}{2.4ex} $\mathcal{S}'$ & $\Lambda_\mathcal{S'} = 100$ & $C_\mathcal{S'} = 0.1$ & -- & $N_{b_{}}^S(0) = 10$  \\
 \rule{0pt}{2ex} $\mathcal{A}'$ & $\Lambda_\mathcal{A'} = 10$ & $C_\mathcal{A'} = 0.1$ & -- & $\alpha = 0.4$  \\
 \rule{0pt}{2ex} $\mathcal{F}$ & $\Lambda_\mathcal{F} = 1$ & $C_\mathcal{F} = 0.1$ & -- & $\beta = 0.04$  \\
 \bottomrule
\end{tabular}
}
\caption{Flow parameters of the {\bf GAI} scenario.\label{tab1}\vspace{-5mm}}
\end{table}
\vspace{-4mm}
\pdfig{8}{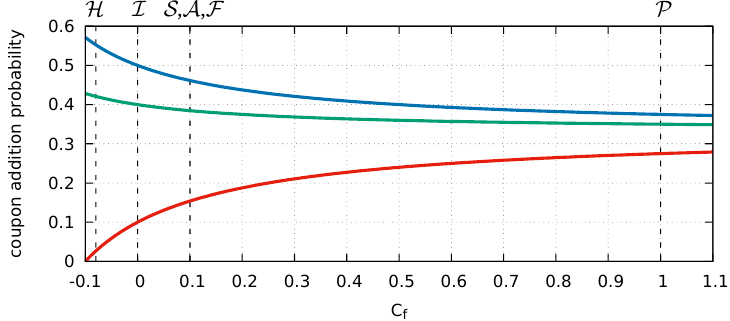}{Impact of $C_f$ on the coupon addition 
probabilities of the red, green and blue primary colors. Vertical dashed lines
correspond to the values chosen for the system flows.}

Before explaining the rationale behind our parameters' choice,
it is useful to observe on Fig. \ref{fig:rplotcropped.pdf} the impact of quality bias parameter $C_f$ on the coupon addition
probabilities $\{r^f_c\}_c$ of each primary component $c$, 
see \,\myeqref{eq:rfi}. Value -0.08, chosen for process $\mathcal{H}$,
reflects the efforts of dataset curators to retain only 
high-quality information. Conversely, the value 1, assigned to 
process $\mathcal{P}$, reflects the limited attention to quality 
for the information initially published on the Web. Moreover, we consider
that the indexing process $\mathcal{I}$ performs a rather good job
at quality discrimination ($C_\mathcal{I} = 0$), while a slightly
less effective filtering (value 0.1) is applied by users on 
machine-generated answers (processes $\mathcal{S},\mathcal{A},\mathcal{F}$).

Table \ref{tab1} reveals a 10:1 ratio between 
rates $\Lambda_\mathcal{S'}$ and $\Lambda_\mathcal{A'}$, reflecting 
the assumption that, within the {\bf GAI} scenario, users still predominantly 
rely on conventional search engines for verifying information 
incorporated into new digital content. We further notice
on Table \ref{tab1}: i) $\gamma = 0.5$, meaning that half of
the information incorporated into the Training Set comes from
online resources; ii) $\mu_S = \lambda_\mathcal{I}$ and
$\mu_L = \lambda_\mathcal{T}$, meaning that the strength
associated to answers learned by both Search Engine and LLM
is discounted over time at the same rate at which it is reinforced.
iii) $\alpha=0.4$, indicating that $40\%$ of answers
generated by GAI tools are assumed to be used in the creation of new content.

Taking this configuration as our reference {\bf GAI} scenario, 
the parameters for the other two scenarios can be readily specified
by difference with respect to those listed in Table \ref{tab1}.
In the {\bf pre-GAI} scenario, we simply set $\Lambda_\mathcal{A} = 0$,
effectively nullifying all effects produced by users employing the 
GAI system.

For the {\bf post-GAI} scenario, we envision a few fundamental parameter shifts
inspired by the current trends in information retrieval practices.   
The primary modification involves an inversion of the rates for processes $\mathcal{S'}$ and $\mathcal{A'}$, with $\Lambda_\mathcal{S'}$ now set to 10 and $\Lambda_\mathcal{A'}$ to 100. In light of the projected 
dominance of GAI-based information retrieval, we establish $\alpha$ at 0.8 and postulate a tenfold increase in the rate of process $\mathcal{H}$ (incorporation of answers in curated datasets), 
setting $\Lambda_\mathcal{H}$ to 1. 
Crucially, we hypothesize that users will increasingly rely on GAI-generated responses, consequently reducing their efforts in quality discrimination when incorporating
answers in new digital content. This shift is reflected in the adjustment of $C_\mathcal{A'}$ from 0.1 (as in the {\bf GAI} scenario) to 1 in the {\bf post-GAI} scenario. 

We commence by presenting select findings from the fundamental RGB model, wherein color mixing is precluded; that is, neither algorithms nor users synthesize novel responses by amalgamating information from primary answers. This model is particularly suitable for queries that elicit some form of discrete, factual information not amenable to synthesis 
from different sources.

\pdfig{7}{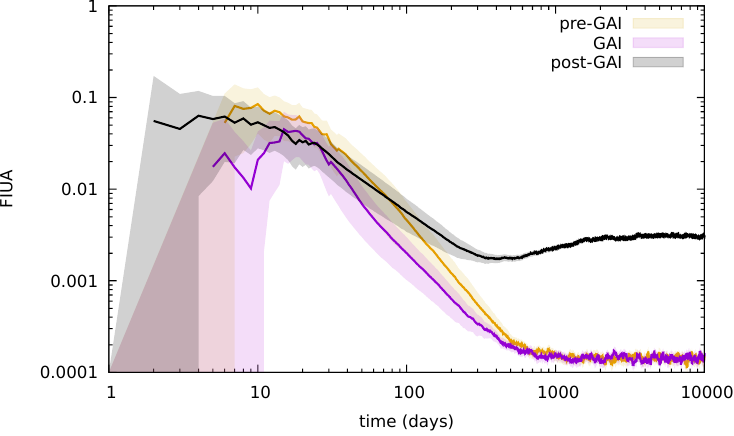}{Temporal evolution of the FIUA indicator for the 
three considered scenarios. 
Shaded areas correspond to 95\%-level
confidence intervals (as in subsequent figures).}

Fig. \ref{fig:FIUA-cropped.pdf} illustrates, as a function of time,  
the fraction of irrelevant used answers (FIUA) found on the Web, under the three
considered scenarios. Shaded regions surrounding each curve represent 95\%-level
confidence intervals derived from 400 simulation runs. These regions reveal an
erratic behavior during the first ten days, after which a clear separation among the curves emerges: interestingly, the {\bf GAI} scenario produces fewer irrelevant answers compared to the {\bf pre-GAI} condition. However, the {\bf post-GAI} scenario ultimately yields the least favorable outcome for this critical metric. Although the reinforcement loop 
of high-quality answers effectively reduces FIUA to relatively low levels across all scenarios, {\it the post-GAI condition generates approximately an order of magnitude more irrelevant answers than the other conditions.}

\begin{figure}[htb]
    \centering
    \begin{subfigure}{0.48\columnwidth}
        \includegraphics[width=\linewidth]{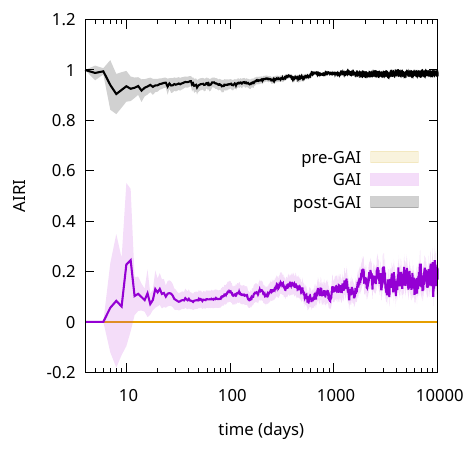}
    \end{subfigure}
    \hfill
    \begin{subfigure}{0.48\columnwidth}
        \includegraphics[width=\linewidth]{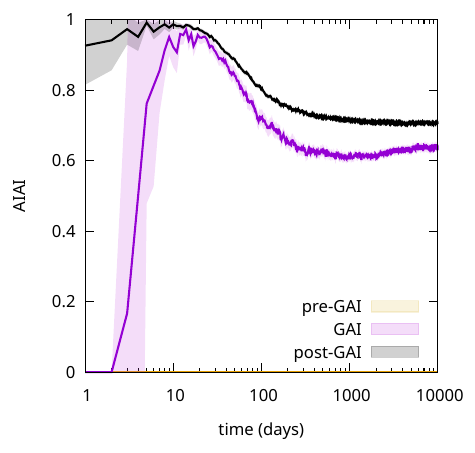}
    \end{subfigure}
    \caption{Temporal evolution of the AIRI (left plot) and
   AIAI indicator (right plot) for the three considered scenarios. 
   }
    \label{fig:combined}
\end{figure}


The AIRI metric (left in Fig. \ref{fig:combined}) reveals
that {\it almost all irrelevant answers produced in the post-GAI scenario come from the generative AI system}.
The AIAI indicator (right in Fig. \ref{fig:combined}) further reveals that {\it the LLM model is affected by a substantial degree of autophagy under both GAI and post-GAI conditions}, ranging from 60\% to 70\%. 

\pdfig{6.5}{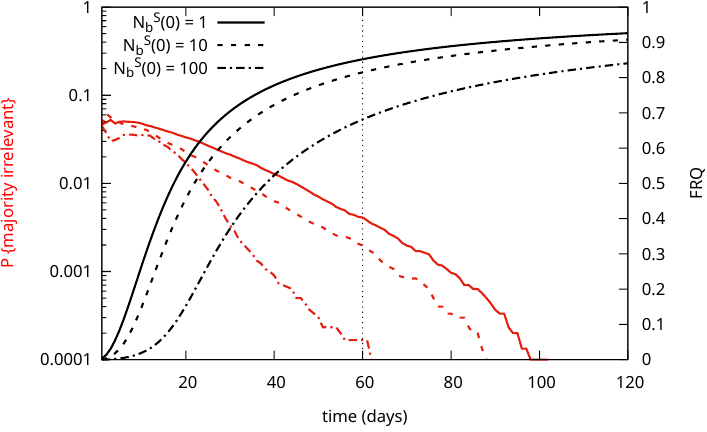}{Probability over time that the majority of used answers 
on the Web are irrelevant (left y-axes), and FRQ indicator (right y-axes) in the 
post-GAI scenario, for initial number of black coupon in the Search Engine
equal to 1,10,100.}

Our system exhibits high stochasticity, particularly during the initial months of topic existence. A notable concern is the potential for irrelevant responses to gain an early, chance-driven advantage over relevant ones. To investigate this critical phenomenon, we conducted 30,000 simulations spanning the first 120 days, calculating the proportion of runs in which, at any given time $t$, irrelevant used
answers in the Web outnumber relevant ones. Focusing on the {\bf post-GAI} scenario, 
we plot this novel metric while varying the initial number $N_b^S(0)$ of black coupons in the Search Engine, which emulates more or less cautious answering strategies. Fig. \ref{fig:fig2-cropped.pdf} presents these results alongside the corresponding Fraction of Responded Queries (FRQ) metric on the secondary 
y-axis. A clear trade-off emerges, whereby more cautious strategies (larger $N_b^S(0)$) can significantly mitigate the probability of the aforementioned critical event, albeit at the expense of diminishing the overall fraction of responded queries
(see for example the values at $t=60$, marked by a vertical dotted line). 
 
\subsection{Results Under Answer Mixing}
This section presents findings from our case study configuration, examining scenarios wherein either the GAI system or users utilizing conventional search engines synthesize novel responses by integrating two distinct information sources. This process progressively populates the system with coupons of intermediate chromatic values, containing various proportions of primary RGB components. 
For a detailed description of the extended model incorporating this fundamental aspect of information synthesis, readers are referred back 
to Sec. \ref{sub:ext}.
Recall that, while we consider the GAI system to uniformely
mix two randomly chosen answers, we instead assume that
users employing the conventional search engine generate
a biased combination of two random answers, applying to the  
best one a weight $\xi \in [1/2,1]$. 
The parameter $\xi$ thus quantifies the average discriminatory 
capacity of users in assessing answer quality. A value of $\xi = 1/2$ 
represents an unbiased mixture, indicating no discrimination, while $\xi = 1$
denotes perfect discrimination, where users consistently 
identify and select the superior information source.

We will separately consider the GAI and post-GAI scenarios,
comparing in each scenario the effects of the following 
conditions: 
\begin{description}[style=unboxed,leftmargin=0.5cm]
\item[no-mix] This corresponds to the base case in which answers
are never mixed. It serves as a baseline for comparative analysis.
\item[mix-GAI] We enable mixing exclusively for the GAI system.
Users employing the Search Engine still use single-source responses.
\item[mix-GAI+SE($\xi$)] We enable mixing for both GAI
and Search Engine. We will investigate the effects of varying the quality discrimination parameter $\xi$, taking values of 0.5, 0.75, and 1.0.
\end{description}  

\pdfig{6.5}{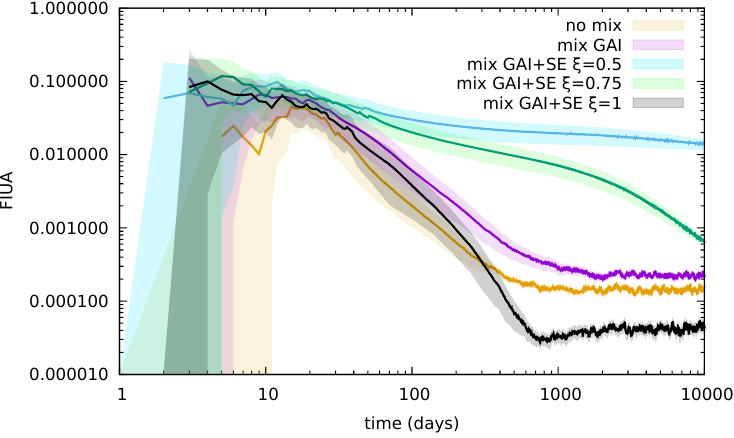}{Temporal evolution of the FIUA indicator 
in the GAI scenario, under different answer mixing assumptions. 
}

Fig. \ref{fig:FIUAmixgpt.pdf} reports the Fraction of Irrelevant Used Answers (FIUA) found in the WWW within the GAI scenario. Recall that here there is a 10:1 ratio
between the utilization rates of answers derived from the SE compared to those 
generated by the GAI. We notice that the mix-GAI condition consistently
yields a higher proportion of irrelevant answers compared to the no-MIX baseline. This observation suggests that the indiscriminate amalgamation of information from diverse sources generally proves detrimental to answer quality.
When we also enable the mixing of SE answers by users, we observe a 
substantial impact of the quality discrimination parameter $\xi$: a markedly
large FIUA is obtained when users indiscriminately combine two sources ($\xi = 0.5$). With $\xi = 0.75$, a very long time is required for the FIUA to approach levels comparable to the no-mix baseline.
Perfect discrimination ($\xi = 1$) is much faster, and ultimately
yields to lowest FIUA among all conditions.

\pdfig{6.5}{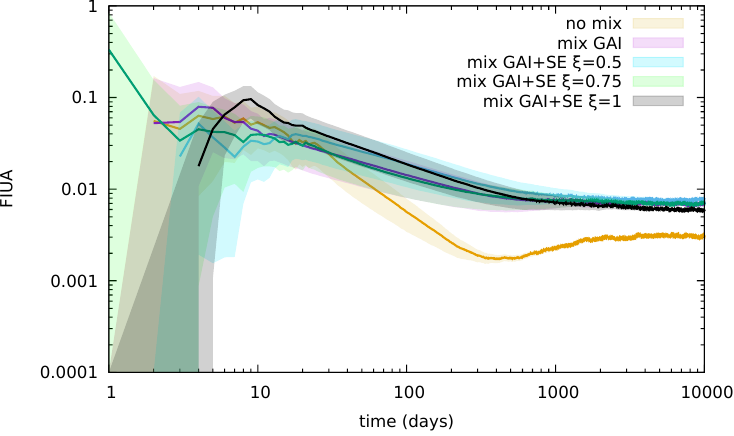}{Temporal evolution of the FIUA indicator 
in the post-GAI scenario, under different answer mixing. 
}


We conducted a similar investigation in the post-GAI scenario. 
Recall that here the vast majority of used answers
come from the GAI system.
The Fraction of Irrelevant Used Answers (FIUA), reported
in Fig. \ref{fig:FIUAmixpost.pdf} confirms that the generation of
responses through the amalgamation of disparate sources consistently yields deleterious effects on the overall quality of information 
across all temporal scales. 
Indeed, the proportion of irrelevant information on the Web 
roughly doubles across
all considered mixing hypotheses, due to the reduced 
impact of search engine-generated responses.
Here the quality discrimination
performed by users has marginal impact because only few users
still employ results by the Search Engine.

%% file: stackoverflow.tex
\begin{figure*}[hbt!]
    \centering
    \includegraphics[width=\textwidth]{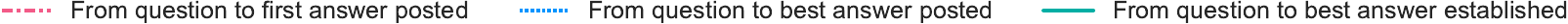}

    \begin{subfigure}[t]{0.45\textwidth}
        \centering
        \includegraphics[width=0.95\textwidth]{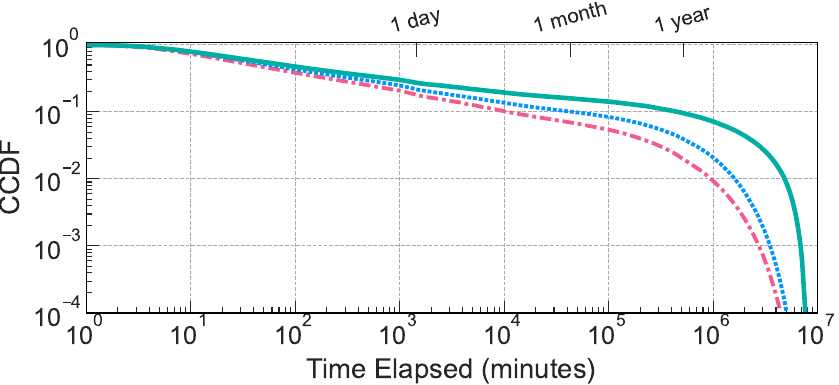}
        \caption{StackOverflow}
        \label{fig:cs_ccdf}
    \end{subfigure}
    \hfill
    \begin{subfigure}[t]{0.45\textwidth}
        \centering
        \includegraphics[width=0.95\textwidth]{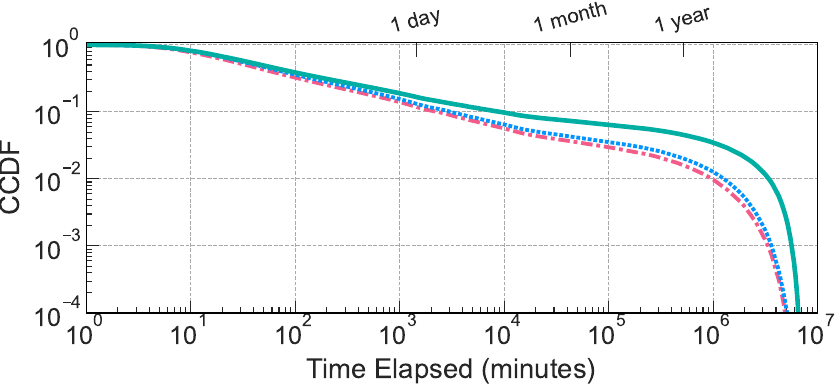}
        \caption{MathStackExchange}
        \label{fig:math_ccdf}
    \end{subfigure}
    \caption{Complementary cumulative distribution function (CCDF) of the time it takes for an answer to be posted (dash-dotted), for the best answer to be posted (dotted), and for the best answer to emerge (solid) (log-log scale).}
    \label{fig:ccdfs}
\end{figure*}
	
\begin{figure}
	\centering
	\includegraphics[width=0.7\columnwidth]{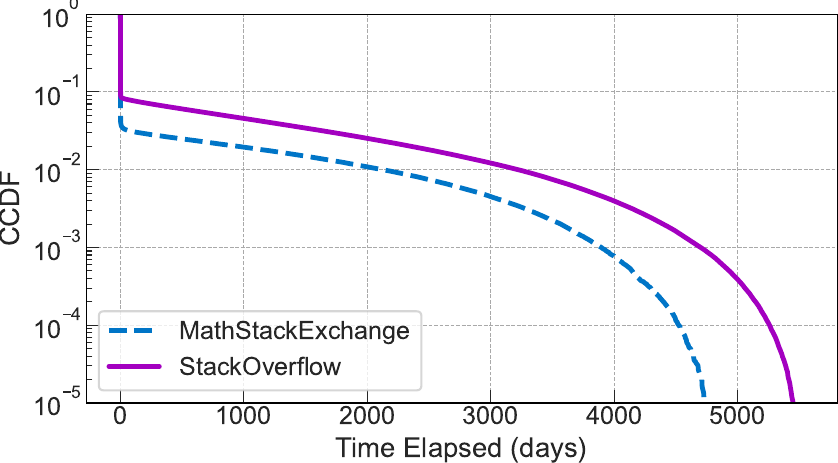}
	\caption{CCDF of the time required, in days, for the best answer to 
be \textit{recognized} as superior for MathStackExchange (dashed) and StackOverflow (solid).} \vspace{-3mm}
	\label{fig:comparison}
\end{figure}

\section{Stack Exchange Analysis}
\label{sec:stack}

Many generative AI systems opt for a chat-based interaction with the user and generally tend to provide a singular, authoritative response even to novel questions.
This precludes the presentation of diverse perspectives and effectively suppresses the competition of ideas that might arise from multiple potential answers. 
To find out to what extent this modality can lead to suboptimal question-and-answer (Q\&A) systems, we investigated one of the most well-known Q\&A platforms: Stack Exchange. In particular, we looked at the largest sub-community, StackOverflow, which mainly deals with computer science and coding, and the largest (non-computer science-related) community, MathStackExchange, which covers a wide range of mathematical topics.

On Stack Exchange, users submit questions, and community members provide answers, which can then be evaluated through upvotes and downvotes. This crowdsourced rating system determines the ranking of answers, with the highest-rated response,
hereafter referred to as \emph{best} answer, displayed first, followed by others in descending order of their accumulated score. 
We extracted~11,555,969 questions from StackOverflow together with 
the corresponding~20,166,328 answers, generated by over 1 million users.
Similarly, we extracted~1,068,196 MathStackExchange questions and their ~1,493,849 answers. The aggregated size of the raw dataset
is about 150 GB. 
We processed this extensive dataset on a HPC cluster, utilizing the Dask Python library to flexibly manage parallel computation. All code utilized in the data processing phase will be made publicly accessible \cite{repo}.



We found that while a large proportion of questions receive the best answer within minutes\footnote{The median time for the \textit{best} answer to appear is 47 min on Math S.E. and 77 min on StackOverflow. The time for the \textit{first} answer to appear is even less: 34 and 39 min.}, a significant subset requires an 
extended period -- sometimes spanning years -- for the optimal response to emerge
(for 10\% of StackOverflow questions, the \textit{best} answer is only recognized after 1 year). 
Figure~\ref{fig:ccdfs} illustrates the complementary cumulative distribution function~(CCDF) of three key time intervals: i) the duration until the 
first answer to a question is submitted (dash-dotted line); ii) the period
until the best answer is posted (dotted line); iii) the interval 
until the best answer achieves its primacy in user display order (solid line).
All intervals above are are calculated from the initial question submission time.
Figure~\ref{fig:comparison} complements previous results by depicting the CCDF of the amount of time required for the best answer to surface, 
measured relative to its initial posting time.

It is evident that there is a significant subset of questions requiring 
an extended temporal interval and significant collective human effort,
both for the generation of high-quality responses and for the community's ultimate recognition of their superiority.